\documentclass[a4paper]{jpconf}
\usepackage{graphicx}
\usepackage{amsmath,amssymb,amsthm,cite,enumerate}

\newcommand{\todo}[1][\null]{\ensuremath{\clubsuit}}
\newcommand{\noprint}[1]{}

\newtheorem{theorem}{Theorem}
\newtheorem{lemma}{Lemma}
\newtheorem{corollary}{Corollary}
\newtheorem{proposition}{Proposition}
{\theoremstyle{definition}
\newtheorem{remark}{Remark}
}

\newcounter{mcasenum}

\begin{document}
\title{Group analysis of  Benjamin--Bona--Mahony equations with time dependent coefficients}

\author{Olena Vaneeva$^1$, Roman  O Popovych$^{1,2}$ and\\[.5ex] Christodoulos Sophocleous$^3$}

\address{$^1$ Institute of Mathematics of NAS of Ukraine, 3 Tereshchenkivska Str., Kyiv 01601, Ukraine}
\address{$^2$ Wolfgang Pauli Institute, Nordbergstra{\ss}e 15, A-1090 Wien, Austria}
\address{$^3$ Department of Mathematics and Statistics, University of Cyprus, Nicosia CY 1678, Cyprus}

\ead{vaneeva@imath.kiev.ua,
rop@imath.kiev.ua,
christod@ucy.ac.cy}

\begin{abstract}
Group classification of a class of Benjamin--Bona--Mahony (BBM) equations with time dependent coefficients is carried out.
Two equivalent lists of equations possessing Lie symmetry extensions are presented:
up to point equivalence within the class of BBM equations and without the simplification by equivalence transformations.
It is shown that the complete results can be achieved using
either the gauging of arbitrary elements of the class by the equivalence transformations or the method of mapping between classes.
As by-product of the second approach the complete group classification of a class of variable-coefficient BBM equations with forcing term is derived.
\end{abstract}

\section{Introduction}

The third-order nonlinear partial differential equation
\begin{equation*}\label{bbm}
u_t+u_x+uu_x-u_{xxt}=0,
\end{equation*}
named these days the Benjamin--Bona--Mahony (BBM) equation, appeared in~\cite{benj1972a}
as an alternative to the Korteweg--de Vries equation, $u_t+u_x+uu_x+u_{xxx}=0$,
model for the unidirectional propagation of moderately long waves with small but finite amplitude in systems that manifest nonlinear and dispersive effects.
Numerical studies showed that the BBM equation admits soliton solutions whose
interaction is inelastic though close to elastic~\cite{abdul1976,bona1983}.
It was proved in~\cite{duzh1984a} for the equivalent form $u_t=uu_x+u_{xxt}$ of the BBM equation that it has no conserved quantity in addition to
those found by Benjamin, Bona and Mahony: $u$ (mass), $(u^2+u_x^2)/2$ (energy), and $u^3/3$ (momentum).
Lie symmetries and the corresponding reductions of the BBM equation in the above equivalent form were obtained in~\cite{kostin1969}
(these results were also presented in~\cite[pp 194--196]{Ibr}).
It was found that the maximal Lie symmetry algebra of this equation
is a~three-dimensional Lie algebra of the type $A_{2.1}\oplus A_1$
spanned by the vector fields $\partial_t$, $t\partial_t-u\partial_u$ and $\partial_x.$

There are several recent works (see~\cite{Molati&Khalique2013} and references therein)
devoted to the study of Lie symmetries of variable-coefficient BBM equations of the general form
\begin{equation}\label{bbm_fgh}
u_t+f(t)u_x+g(t)uu_x+h(t)u_{xxt}=0,
\end{equation}
where $f$, $g$ and $h$ are arbitrary smooth functions of the variable $t$ with $gh\neq0$.
However none of these works contains exhaustive and completely correct results.
We aim to fill up this gap by presenting the exhaustive group classification of equations from class~\eqref{bbm_fgh}
and classifying local conservation laws of these equations.

In order to study transformational properties of class~\eqref{bbm_fgh},
in Section~2 we describe the entire set of its admissible transformations (i.e., the equivalence groupoid of this class)
by proving that the class is normalized in the usual sense.
The assertions concerning reducibility of variable-coefficient BBM equations to their constant-coefficient counterparts are presented therein.
The complete classification of Lie symmetries is obtained in Section~3
under gauging the arbitrary element~$g$ to~1 by a parameterized family of point transformations
related to the associated equivalence group.
The method of mapping between classes is utilized in Section~4 to verify the classification results.
As by-product of this approach
the complete group classification of a~related class of variable-coefficient BBM equations with forcing term, $u_t+ uu_x+h(t)u_{xxt}=l(t)$, is derived.
In Section~5 we construct local conservation laws of equations from the class~\eqref{bbm_fgh}.

\section{Admissible transformations}

Two differential equations which are connected by a point transformation are called {\it similar}~\cite{Ovsiannikov1982}.
Similar differential equations have similar sets of solutions, symmetries, conservation laws and other related objects.
Therefore, it is instructive to study point transformations linking equations from a given class of differential equations.
Such transformations are called form-preserving~\cite{Kingston&Sophocleous1998} or allowed~\cite{Winternitz92}.
The rigorous reformulation  of this notion leads to the notions of admissible transformations~\cite{popo2006b,popo2010a}
and the equivalence groupoid of the class \cite{Popovych&Bihlo2012,BihloPopovychJMP2012}.
An admissible transformation is a triple that consists of two similar equations from the class and a~point transformation between them.
The set of admissible transformations of the class possesses the groupoid structure
with respect to the natural operation of the composition of admissible transformations
and hence it is called the \emph{equivalence groupoid} of the class.

A joint transformation of the unknown functions and the arbitrary elements is called an equivalence transformation of the class
if it satisfies the following properties:
(\emph{i}) The transformation becomes point with respect to the independent and the dependent variables when the arbitrary elements are fixed.
(\emph{ii}) It is consistent with the contact structure of the underlying jet space.
(\emph{iii}) It maps any equation from the class to an equation from the same class.
Depending on additional restrictions, one have different kinds of equivalence transformations
(usual, generalized, extended and generalized extended ones).
Equivalence transformations of the same kind constitute a group, which is called the equivalence group of the class (of the respective kind),
see \cite{popo2010a,kpv2014} and references therein.
If the equivalence groupoid of the class is generated by the equivalence group of this class,
then this class is  called {\it normalized} in the same sense as the kind of the equivalence group involved~\cite{popo2006b,popo2010a}.

One-parameter subgroups of the equivalence group can be found by the Lie infinitesimal method,
whereas the direct method~\cite{Kingston&Sophocleous1991,king1991a,Kingston&Sophocleous1998} allows one
to construct the entire equivalence group including discrete equivalence transformations and this technique is preferable.
The direct method can also be used for finding the equivalence groupoid and checking whether the class is normalized
 \cite{popo2010a,Popovych&Bihlo2012,BihloPopovychJMP2012,vane2014ppv}.
A very useful feature of normalized classes is that the equivalence groups
for their subclasses, singled out by setting additional constraints on arbitrary elements,
are subgroups of the equivalence group of the entire class.
We use this property to derive the equivalence group of the class~\eqref{bbm_fgh} from the equivalence group of a normalized superclass.

Consider the class of (1+1)-dimensional partial differential equations of  order $p+q>1$ with fixed $p,q\in\mathbb{N}_0,$
\begin{equation}\label{eq_genClass}
u_{pq}=H\left(t,x,u_{(p+q-1)}\right),
\end{equation}
where $u_{pq}=\partial^{p+q}u/\partial t^p\partial x^q$ and $u_{(p+q-1)}$ denotes the set of all derivatives of $u$  up to order $p+q-1$ including $u$ as the zero-order derivative.
The function $H$ is an arbitrary smooth function of its arguments. The following assertion was proven in~\cite{Kingston&Sophocleous1998}.

\begin{theorem}
A point transformation
$\tilde t=T(t,x,u)$, $\tilde x=X(t,x,u)$, $\tilde u=U(t,x,u)$,
where $\left|\partial(T,X,U)/\partial(t,x,u)\right|\ne0$,
maps a fixed equation from class~\eqref{eq_genClass} to an equation from the same class
if and only if it satisfies the following conditions:
\begin{enumerate}
  \item[$({\it i})$] $T=T(t)$, $X=X(x)$\quad for\quad $pq\neq0$,\quad $p\neq q$;
  \item[$({\it ii})$] $T=T(t)$, $X=X(x)$\quad or\quad $T=T(x)$, $X=X(t)$\quad for\quad $p=q\ne0$;
  \item[$({\it iii})$] $T=T(t)$\quad for\quad $p=0$, $q\neq0$;
  \item[$({\it iv})$] $X=X(x)$\quad for\quad $p\neq0$, $q=0$.
\end{enumerate}
\end{theorem}

\begin{corollary}
The usual equivalence group of class~\eqref{eq_genClass} with any fixed $p$ and $q$ consists of the transformations whose projections to the space of~$(t,x,u)$
coincides with that described in Theorem~1. Moreover, this class is normalized in the usual sense.
\end{corollary}

Therefore, Theorem~1 provides the complete description of admissible (form-preserving) transformations in class~\eqref{eq_genClass}.
As this class is normalized, we can use transformation constraints from Theorem~1 for describing admissible transformations in subclasses of this class,
in particular in class \eqref{bbm_fgh}.

For class~\eqref{bbm_fgh} we have $p=1$ and $q=2$, which corresponds to Case~$({\it i})$ of Theorem~1.
Now we directly seek for all point transformations of the general form
\begin{equation}\label{eq:GeneralEquivalenceTransformation}
    \tilde t = T(t), \quad \tilde x = X(x), \quad \tilde u = U(t,x,u)
\end{equation}
with  $T_tX_xU_u\neq0$  that map a fixed equation of the form~\eqref{bbm_fgh} to an equation from the same class,
\begin{equation}\label{eqBBMtilded}
    \tilde u_{\tilde t}+\tilde f(\tilde t)\tilde u_{\tilde x}+\tilde g(\tilde t)\tilde u\tilde u_{\tilde x}+\tilde h(\tilde t)\tilde u_{\tilde x\tilde x\tilde t}=0.
\end{equation}
The components of the prolongation of transformation~\eqref{eq:GeneralEquivalenceTransformation} to the involved partial derivatives are
\begin{gather*}
\tilde u_{\tilde t}=\frac1{T_t}\left(U_t+U_uu_t\right),\quad \tilde u_{\tilde x}=\frac1{X_x}\left(U_x+U_uu_x\right),\\\tilde u_{\tilde x\tilde x\tilde t}=-\frac{X_{xx}}{T_tX_x^3}\bigl(U_{tx}+U_{xu}u_t+U_{tu}u_x+U_{uu}u_tu_x+U_uu_{tx}\bigr)+{}\\
{}+\frac1{T_tX_x^2}\bigl( U_{xxt}+U_{xxu}u_t+2U_{txu}u_x+2U_{xuu}u_tu_x+2U_{xu}u_{tx}+U_{tuu}u_x^2+U_{uuu}u_tu_x^2+{}\\
{}+2U_{uu}u_xu_{tx}+U_{tu}u_{xx}+U_{uu}u_tu_{xx}+U_uu_{xxt}\bigr).
\end{gather*}
Using these expressions, we rewrite equation~\eqref{eqBBMtilded} in the old variables, which are without tildes.
Then we substitute $u_{xxt}=-(u_t+f(t)u_x+g(t)uu_x)/h(t)$ into the obtained equation
in order to confine it to the manifold defined by~\eqref{bbm_fgh} in the third-order jet space
with the independent variables $(t,x)$ and the dependent variable~$u$.
Splitting the resulting identity with respect to the derivatives $u_t$, $u_x$, $u_{tx}$ and $u_{xx}$ leads to the determining
equations on the transformation components~$T$, $X$, and~$U$,
\begin{gather*}
U_{uu}=U_{tu}=0,\quad X_{xx}U_u=2X_x U_{xu},\quad X_x^3U_u=\tilde h(X_{xx}U_{xu}-X_xU_{xxu})+\frac{\tilde h}hX_xU_u,\\
U_t+\frac{T_t}{X_x}(\tilde f+\tilde gU)U_x={\tilde h}\frac{X_{xx}}{X_x^3}U_{tx}-{\tilde h}\frac1{X_x^2}U_{xxt},\quad
\tilde f+\tilde gU=\dfrac{\tilde h}h\dfrac{f+gu}{T_tX_x}.
\end{gather*}
The equations $U_{uu}=U_{tu}=0$ imply the representation $U=U^1(x)u+U^0(t,x)$,
where~$U^1$ and~$U^0$ are smooth functions of their arguments with $U^1\neq0.$
We substitute this representation for~$U$ into the remaining  determining equations and further split with respect to $u$.
As a result, we get the equations $U^1_x=U^0_t=U^1_x=X_{xx}=0,$
$\tilde f=\dfrac{X_x}{T_tU^1}(U^1 f-U^0 g),$ $\tilde g=\dfrac{X_x}{T_tU^1}g,$ and $\tilde h=X_x^2h.$
We solve the equations for $X$, $U^1$ and $U^0$, and substitute the obtained expressions
into the equations representing transformation components for arbitrary elements, which leads to the following assertion.

\begin{theorem}
The usual equivalence group~$G^{\sim}_1$ of class~\eqref{bbm_fgh} is comprised of the transformations
\[
\begin{array}{c}
\tilde t=T(t),\quad \tilde x=\delta_1x+\delta_2,  \quad
\tilde u=\delta_3u+\delta_4, \quad
\tilde f=\dfrac{\delta_1}{T_t\delta_3}(\delta_3 f-\delta_4g), \quad
\tilde g=\dfrac{\delta_1}{T_t\delta_3}g,\quad
\tilde h=\delta_1^2h,
\end{array}
\]
where $\delta_j$, $j=1,2,3,4$, are arbitrary constants with $\delta_1\delta_3\not=0$
and $T=T(t)$ is an arbitrary smooth function with $T_t\neq0.$
Class~\eqref{bbm_fgh} is normalized in the usual sense.
\end{theorem}

Thus, each point transformation between equations from class~\eqref{bbm_fgh} is induced by an element of the group~\smash{$G^{\sim}_1$}.
In order to find which variable-coefficient equations of the form~\eqref{bbm_fgh} admit constant-coefficient counterparts,
we assume that the transformed arbitrary elements~$\tilde f$, $\tilde g$ and~$\tilde h$ are constants in equivalence transformations.
This results in the following assertion:

\begin{proposition}
A variable-coefficient equation from class~\eqref{bbm_fgh} is reduced to a constant-coefficient equation from the same class  by a point transformation
if and only if
the corresponding coefficients $f,$ $g$ and $h$ satisfy the conditions
\begin{equation*}\label{criterion1}
\left(\frac fg\right)_t=h_t=0,
\end{equation*}
i.e., $h$ is a constant and $f$ is proportional to $g$.
\end{proposition}

Equivalence transformations allow us to simplify the consideration by reducing the number of arbitrary elements.
For example, we can set the gauge $g=1$ using the family of point transformations
\begin{equation}\label{tr}
\tilde t=\int\!g(t){\rm d}t,\quad\tilde x=x,\quad\tilde u=u
\end{equation}
parameterized by the arbitrary element~$g$ and related to equivalence transformations from the group~$G^{\sim}_1$.
Then the other arbitrary elements are $\tilde f(\tilde t)=f(t)/g(t),$ and $\tilde h(\tilde t)=h(t)$.
Here and below an integral with respect to $t$ should be interpreted as a fixed antiderivative.

Therefore, without loss of generality we can restrict ourselves by the study of the class
\begin{equation}\label{bbm_fh}
u_t+f(t)u_x+uu_x+h(t)u_{xxt}=0,
\end{equation}
as all results on symmetries and exact solutions for equations of the form~\eqref{bbm_fgh} can be recovered
from the corresponding results obtained for equations of the form~\eqref{bbm_fh} using the above family of point transformations.

Since class~\eqref{bbm_fgh} is normalized, the equivalence group of its subclass~\eqref{bbm_fh} can be easily found
as the subgroup of the group~$G^{\sim}_1$ whose elements preserve the gauge~$g=1$.

\begin{corollary}
Class~\eqref{bbm_fh} is normalized in the usual sense.
Its usual equivalence group~$G^{\sim}_2$ is constituted by the transformations
\[
\begin{array}{l}
\tilde t=\dfrac{\delta_1}{\delta_3}t+\delta_0,\quad \tilde x=\delta_1x+\delta_2,  \quad
\tilde u=\delta_3u+\delta_4, \quad
\tilde f=\delta_3 f-\delta_4, \quad
\tilde h=\delta_1^2h,
\end{array}
\]
where $\delta_j,$ $j=0,\dots,4, $ are arbitrary constants with $\delta_1\delta_3\not=0$.
\end{corollary}

There are no truly variable-coefficient equations in class~\eqref{bbm_fh} that are reduced by point transformations to constant-coefficient equations from the same class.

In the next section we use the equivalence transformations in the course of group classification of classes~\eqref{bbm_fgh} and~\eqref{bbm_fh}.

\section{Lie symmetries}

We carry out the group classification of class~\eqref{bbm_fh} within the framework of the classical Lie approach~\cite{Olver1993,Ovsiannikov1982}.
We look for vector fields of the form
\[
\Gamma=\tau(t,x,u)\partial_t+\xi(t,x,u)\partial_x+\eta(t,x,u)\partial_u
\]
which generate one-parameter groups of point-symmetry transformations of an equation from class~\eqref{bbm_fh}
and hence jointly constitute the maximal Lie invariance algebra~$A^{\max}$ of this equation.
Any such vector field,~$\Gamma$, satisfies the infinitesimal invariance criterion, i.e.,
the action of the third prolongation,~$\Gamma^{(3)}$, of the vector field~$\Gamma$ on equation~\eqref{bbm_fh}
results in the conditions being an identity for all solutions of this equation.
That is, the criterion is read as
\begin{gather}\label{conditionf(t)3}\arraycolsep=0ex
\begin{array}{l}
\Gamma^{(3)}\left[u_t+f(t)u_x+uu_x+h(t)u_{xxt}
\right]\big|_{u_{xxt}=-(u_t+f(t)u_x+uu_x)/h(t)}=0.
\end{array}
\end{gather}

Theorem~2 allows us to simplify the computation since it implies that
$\tau=\tau(t)$, $\xi=\xi(x)$, $\eta=\eta(u)$, and $\xi_{xx}=\eta_{uu}=0$,
Then the left hand side of the condition~\eqref{conditionf(t)3} is a polynomial in the variables~$x$, $u$, $u_t$ and~$u_x$
with coefficients depending on~$t$.
Hence these coefficients are equal to zero,
which gives the determining equations on remaining arbitrarinesses in the components~$\tau$, $\xi$ and~$\eta$ of the vector field~$\Gamma$.
Solving the simplest of these equations we get the general form of the components of a Lie symmetry operator of an equation from class~\eqref{bbm_fh},
\begin{equation*}
\tau=\left(\tfrac12c_2-c_1\right) t+c_0,\quad \xi=c_2x+c_3,\quad \eta=\left(\tfrac12c_2+c_1\right) u+c_4
\end{equation*}
with $c_i$, $i=0,\dots,4,$ being arbitrary constants.
The residuary determining equations involving arbitrary elements have the form
\begin{gather}\label{c1}
\left(\left(\tfrac12c_2-c_1\right)t+c_0\right)\!h_t=2c_2h,\quad
\left(\left(\tfrac12c_2-c_1\right)t+c_0\right)\!f_t=\left(\tfrac12c_2+c_1\right)\!f-c_4
\end{gather}
and are called {\it classifying equations}.
Further analysis is carried out using the {\it method of furcate splitting} introduced in~\cite{Nikitin&Popovych2001}.
For each vector field $\Gamma$ from the algebra~$A^{\max}$ equations~\eqref{c1}
gives a~system on~$f$ and $h$ of the general form
\begin{equation}\label{classif_eq}
(\alpha\,t+\beta)h_t=\gamma h,\quad (\alpha\,t+\beta)f_t=\big(\tfrac12{\gamma}-\alpha\big) f+\nu,
\end{equation}
where $\alpha$, $\beta$, $\gamma$ and $\nu$ are constants.
Varying vector fields from $A^{\max}$ we obtain a set of systems of the form~\eqref{classif_eq}.
The number $k$ of such systems with linearly independent tuples of the coefficients $\alpha$, $\beta$, $\gamma$ and $\nu$
is not greater than two; otherwise the set of these systems would be inconsistent in total.
Therefore, possible values of $k$ are $k=0$, $k=1$ and $k=2$.

If $k=0$, then equations~\eqref{c1} are identities with respect to~$f$ and~$h$  and hence $c_0=c_1=c_2=c_4=0$.
Thus, we obtain that the maximal Lie invariance algebras of equations from class~\eqref{bbm_fh} for general values of $f$ and $h$
is the one-dimensional algebra $\langle\partial_x\rangle$, which gives Case~0 of Table~1.
We get the same result when varying the arbitrary elements~$f$ and~$h$ an splitting equations~\eqref{c1} with respect to them.
In other words, the kernel of the maximal Lie invariance algebras of equations from class~\eqref{bbm_fh}
(i.e., the common part of the maximal Lie invariance algebras of equations from this class, see~\cite{Ovsiannikov1982} for the definition)
coincides with the maximal Lie invariance algebra $\langle\partial_x\rangle$ of the general case.

If $k=1$, equations~\eqref{classif_eq} are not identities but equations on $f$ and $h$.
In order to get $G^{\sim}_2$-inequivalent solutions of these equations,
we study how transformations from $G^{\sim}_2$ act on the coefficients $\alpha$, $\beta$, $\gamma$ and $\nu$ of system~\eqref{classif_eq}.
Combined with the multiplication by a nonzero constant~$\chi$,
each transformation from the equivalence group~$G^{\sim}_2$ is extended to the coefficient quadruple
of the system~\eqref{classif_eq} as
\begin{gather*}
\tilde\alpha=\alpha{\chi},\quad\tilde\beta=\left(\frac{\delta_1}{\delta_3}\beta-\delta_0\alpha\right)\!{\chi},
\quad\tilde\gamma=\gamma{\chi},\quad\tilde\nu=\left({\delta_3}\nu+{\delta_4}\left(\tfrac12{\gamma}-\alpha\right)\right){\chi}.
\end{gather*}
Analysis of the induced transformations for $\alpha$, $\beta$, $\gamma$ and $\nu$ results in the following assertion.

\begin{lemma}\label{LemmaOnTransOfCoeffsOfClassifyingSystem2}
Up to $G^{\sim}_2$-equivalence
the parameter quadruple~$(\alpha,\beta,\gamma,\nu)$ can be assumed to belong to the set
\[
\{(1,0,\rho,0),\ (1,0,2,1),\ (0,1,1,0),\ (0,1,0,1)\},
\]
where $\rho$ is an arbitrary constant.
\end{lemma}

The group classification  for the case $k=1$ is obtained as follows.
Sequently solving system~\eqref{classif_eq} for each of the four inequivalent values of $(\alpha,\beta,\gamma,\nu)$ presented in Lemma~\ref{LemmaOnTransOfCoeffsOfClassifyingSystem2},
we get the corresponding cases for the arbitrary elements $(h,f)$,
\[
\big(\varepsilon t^\rho,\lambda t^{\frac{\rho-2}2}\big),\quad
(\varepsilon t^2,\ln t+\lambda),\quad
\big(\varepsilon e^t,\lambda e^{\frac{t}2}\big),\quad
(\varepsilon,t+\lambda),
\]
where $\varepsilon$ and $\lambda$ are constant parameters arising in the course of integration, and $\varepsilon\neq0.$
Then we check whether the remaining parameters can be additionally gauged by transformations from the group~$G^\sim_2$.
It appears that~$\varepsilon$ can be scaled to~$1$ if it is positive and to~$-1$ if it is negative.
In~the second and the fourth cases the constant $\lambda$ can be set equal to zero by a translation of~$u$.
In the first and the third cases~$\lambda$ cannot be gauged.
We substitute the obtained forms of~$h$ and~$f$ into equations~\eqref{c1} to get constraints on constants $c_i$, $i=0,\dots,4$.
This gives the respective forms of Lie symmetry operators. The results are collected in Cases~1--4 of Table~1.%
\footnote{%
Arguments of logarithm and bases in powers with noninteger exponents containing $\rho$ should be carefully treated throughout the paper.
In fact, one should take their absolute values if they can be negative.
Usually the positivity of~$t$ well agrees with the physical meaning of $t$ as time variable.
So, omitting the notation of absolute value is not too essential in Table~1.
At the same time, the expression $\varepsilon T+\kappa$ from Table~2 may be negative and hence we implicitly assume replacing it by its absolute value.
}

If $k=2$, then up to $G^\sim_2$-equivalence and linear combining there exists a single pair of the coefficient tuples,
namely, $(1,0,0,1)$ and $(0,1,0,0)$, for which the corresponding systems of the form~\eqref{classif_eq} are consistent to each other.
Then $f$ and $h$ are constants, $f=0\bmod G^\sim_2$ and $h=\varepsilon\bmod G^\sim_2$, where $\varepsilon=\pm1$,
which gives Case~5 of Table~1.

We sum up the above consideration.

\begin{theorem}
The kernel of the maximal Lie invariance algebras of equations from class~\eqref{bbm_fh}
is the one-dimensional  algebra $\langle\partial_x\rangle$.
All possible $G^\sim_2$-inequivalent cases of Lie symmetry extensions are exhausted by Cases 1--5 of Table~1.
\end{theorem}

\begin{corollary}
The group classification for class~\eqref{bbm_fgh} up to $G^\sim_1$-equivalence results in the list presented in Table~1,
where the arbitrary element~$g$ is assumed to be equal~$1$.
\end{corollary}

\begin{table}[h!]
\caption{\label{TableLieSym1}The group classification of class~\eqref{bbm_fh} up to $G^\sim_2$-equivalence.
}
\begin{center}
\begin{tabular}{cccl}
\br
no.&$h(t)$&$f(t)$&\hfil Basis of $A^{\max}$ \\
\mr
0&$\forall$&$\forall$&$\partial_x$\\
1&$\varepsilon t^\rho$&$\lambda t^{\frac{\rho-2}2}$&
$\partial_x,\ 2t\partial_t+\rho x\partial_x+(\rho-2)u\partial_u$\\
2&$\varepsilon t^2$&$\ln t$&
$\partial_x,\  t\partial_t+x\partial_x-\partial_u$\\
3&$\varepsilon e^t$&$\lambda e^{\frac12t}$&
$\partial_x,\  2\partial_t+x\partial_x+u\partial_u$\\
4&$\varepsilon$&$t$&
$\partial_x,\ \partial_t-\partial_u$\\
5&
$\varepsilon$&$0$&$\partial_x,\ \partial_t,\  t\partial_t-u\partial_u$\\
\br
\end{tabular}
\\[1ex]
\parbox{140mm}{Here $\rho$ and $\lambda$ are arbitrary constants, $\varepsilon=\pm1\bmod G^\sim_2$. In Case 1 $(\rho,\lambda)\neq(0,0)$.}\\[1ex]
\caption{\label{TableLieSym3}The group classification of class~\eqref{bbm_fgh}
using no equivalence.}
\begin{tabular}{cccl}
\br
no.&$h(t)$&$f(t)$&\hfil Basis of $A^{\max}$ \\
\mr
0&$\forall$&$\forall$&$\partial_x$
\\\rule{0ex}{4.2ex}
1&$\mu_1(\varepsilon T+\kappa)^\rho$&$\mu_2g(\varepsilon T+\kappa)^{\frac{\rho-2}2}+\mu_3g$&
$\partial_x,\ \dfrac2g(\varepsilon T+\kappa)\partial_t+\varepsilon\rho x\partial_x+\varepsilon(\rho-2)(u+\mu_3)\partial_u$
\\\rule{0ex}{4.2ex}
2&$\mu_1(\varepsilon T+\kappa)^2$&$\mu_2g\ln(\varepsilon T+\kappa)+\mu_3g$&
$\partial_x,\  \dfrac1g(\varepsilon T+\kappa)\partial_t+\varepsilon x\partial_x-\varepsilon\mu_2\partial_u$
\\\rule{0ex}{4.2ex}
3&$\mu_1 \exp({\sigma T})$&$\mu_2 g\exp({\frac12\sigma T})+\mu_3g$&
$\partial_x,\  \dfrac2g\partial_t+\sigma x\partial_x+\sigma(u+\mu_3)\partial_u$
\\\rule{0ex}{4.2ex}
4&$\mu_1$&$\mu_2gT+\mu_3g$&$\partial_x,\ \dfrac1g\partial_t-\mu_2\partial_u$
\\\rule{0ex}{4.2ex}
5&$\mu_1$&$\mu_3g$&$\partial_x,\ \dfrac1g\partial_t,\  \dfrac Tg\partial_t-(u+\mu_3)\partial_u$
\\
\br
\end{tabular}
\\[1ex]
\parbox{158mm}{Here $g$ is an arbitrary nonvanishing smooth function, $T=\int\!g(t)\,{\rm d}t$; $\varepsilon=\pm1$;
$\mu_1$, $\mu_2$, $\mu_3$, $\nu$ and~$\rho$ are arbitrary constants satisfying the following constraints:
$\mu_1\lambda\neq0$; in Case~1 $\rho\mu_2\neq0$; and in Case~4 $\mu_2\neq0$.}
\end{center}
\end{table}

In order to get the classification list for class~\eqref{bbm_fgh}, where forms of arbitrary elements are not simplified by equivalence transformations, we apply transformation~\eqref{tr} combined with transformations from the equivalence group $G^\sim_2$ to equations of the form~\eqref{bbm_fh} with $f$ and $h$ presented in Table~1.
Basis elements of the corresponding maximal Lie invariance algebras are pushed forward by the same transformations.
Then we re-denote the constants and collect the obtained results in Table~2.
The detailed procedure of the {\it equivalence based approach} for deriving most general forms of arbitrary elements and basis elements of the corresponding maximal Lie invariance algebras can be found in~\cite{Vaneeva2012}.

\section{Mapping between classes and group classification of the related class}

An alternative way for group classification of class~\eqref{bbm_fgh}
is the method of mapping between classes, which was suggested in~\cite{VPS2009}.
This method has been successfully applied to several classes of nonlinear partial differential equations (see, e.g.,~\cite{Vaneeva&Kuriksha&Sophocleous2015}).

Class~\eqref{bbm_fgh}  can be mapped to a similar class of third-order partial differential equations of  the form
\begin{equation}\label{eqBBMimaged}
u_t+ uu_x+h(t)u_{xxt}=l(t),\quad h\neq0.
\end{equation}
The map is realized by the family of point transformations
\begin{equation}\label{trr}
\tilde t=\int\!\! g(t){\rm d}t,\quad \tilde x=x,\quad\tilde u=u+\frac{{f}(t)}{g(t)},
\end{equation}
parameterized by two arbitrary elements of class~\eqref{bbm_fgh}.
The arbitrary elements in the imaged equations take values (tildes in~\eqref{eqBBMimaged} are omitted)
\[\tilde h(\tilde t)=h(t),\quad l(\tilde t)=\frac1{g(t)}\left(\frac{f(t)}{g(t)}\right)_t.\]
Following the method of mapping between classes, we first classify Lie symmetries of the imaged class~\eqref{eqBBMimaged}
and then use the family of point transformations~\eqref{trr} to extend the result to the initial class~\eqref{bbm_fgh}.

In order to efficiently solve the group classification problem for class~\eqref{eqBBMimaged},
we look for admissible transformations in this class using the direct method.
It appears that such transformations are exhausted by transformations from the usual equivalence group of this class.

\begin{theorem}
The usual equivalence group~$G^{\sim}_3$ of class~\eqref{eqBBMimaged} consists of the transformations
\[
\begin{array}{c}
\tilde t=\dfrac{\delta_1}{\delta_3}t+\delta_0,\quad \tilde x=\delta_1x+\delta_2,  \quad
\tilde u=\delta_3u,\quad
\tilde h=\delta_1^2h,\quad
\tilde l=\dfrac{\delta_3^2}{\delta_1}l,
\end{array}
\]
where $\delta_j$, $j=0,1,2,3$, are arbitrary constants with $\delta_1\delta_3\not=0$.
Class~\eqref{eqBBMimaged} is normalized in the usual sense.
\end{theorem}

Using the classical Lie infinitesimal method in the same way as in the previous section,
we get the complete group classification of equations from class~\eqref{eqBBMimaged}.
The results are summarized in the following assertion.

\begin{theorem}
The kernel of the maximal Lie invariance algebras of equations from class~\eqref{eqBBMimaged}
is the one-dimensional  algebra $\langle\partial_x\rangle$.
All possible $G^\sim_3$-inequiva\-lent  cases of extension of the maximal Lie invariance algebras are exhausted
by Cases $1$--$\,4$ of Table~\ref{TableLieSym4}.
\end{theorem}

\begin{table}[ht]
\caption{\label{TableLieSym4}The group classification of  class~\eqref{eqBBMimaged} up to $G^\sim_3$-equivalence.}
\begin{center}
\begin{tabular}{cccl}
\br
no.&$h(t)$&$l(t)$&\hfil Basis of $A^{\max}$ \\
\mr
0&$\forall$&$\forall$&$\partial_x$\\
1&$\varepsilon t^\rho$&$\lambda t^{\frac{\rho-4}{2}}$&
$\partial_x,\ 2t\partial_t+\rho x\partial_x+(\rho-2)u\partial_u$\\
2&$\varepsilon e^t$&$\lambda e^{\frac12t}$&
$\partial_x,\  2\partial_t+x\partial_x+u\partial_u$\\
3&$\varepsilon$&$1$&
$\partial_x,\ \partial_t$\\
4&
$\varepsilon$&$0$&$\partial_x,\ \partial_t,\  t\partial_t-u\partial_u$\\
\br
\end{tabular}
\\[1ex]
\parbox{140mm}{Here $\rho$ and $\lambda$ are arbitrary constants, $\varepsilon=\pm1\bmod G^\sim_3,$ in Case 1 $(\rho,\lambda)\neq(0,0)$. }
\end{center}\end{table}
\begin{remark}
The most general forms of the functions $h$ and $l$ that correspond to equations from class~\eqref{eqBBMimaged} with Lie symmetry extensions are

\smallskip

1. $h=\lambda_1 (\varepsilon t+\kappa)^\rho$, $l=\lambda_2 (\varepsilon t+\kappa)^{\frac{\rho-4}{2}}$:\quad
$A^{\rm max}=\left\langle\partial_x,\,2(\varepsilon t+\kappa)\partial_t+\varepsilon\rho x\partial_x+\varepsilon(\rho-2)u\partial_u\right\rangle$;

\smallskip

2. $h=\lambda_1 e^{\sigma t}$, $l=\lambda_2 e^{\frac 12\sigma t}$:\quad
$A^{\rm max}=\langle\partial_x,\,2\partial_t+\sigma x\partial_x+\sigma u\partial_u\rangle$;

\smallskip

3.  $h=\lambda_1 $, $l=\lambda_2$:\quad
$A^{\rm max}=\langle\partial_x,\,\partial_t\rangle$;

\smallskip

4.  $h=\lambda_1 $, $l=0$:\quad
$A^{\rm max}=\langle\partial_x,\,\partial_t,\, t\partial_t-u\partial_u\rangle$.

\smallskip

\noindent
Here $\lambda_1$, $\lambda_2$, $\varepsilon$, $\kappa$ and $\rho$ are arbitrary constants with $\lambda_1\sigma\varepsilon\neq0$.
Additionally, in Case~1 $(\rho,\lambda_2)\neq(0,0)$ and in Case~3 $\lambda_2\neq0$.
Due to the presence of arbitrary constants~$\lambda_1$ and~$\lambda_2$, the constant~$\varepsilon$ can be assumed to take the values~$\pm1$ only.
\end{remark}

The following example shows how to recover the group classification of class~\eqref{bbm} using the results obtained for class~\eqref{eqBBMimaged}.

Consider Case~1 of Table~3 extended by the equivalence transformations from~$G^\sim_3$, i.e., the first case presented in Remark~3,
were $\tilde h=\lambda_1 (\varepsilon \tilde t+\kappa)^\rho$, $l=\lambda_2(\varepsilon \tilde t+\kappa)^\frac{\rho-4}{2}.$
We denote $\int\! g(t)\, {\rm d}t$ by $T$.
As $\tilde t=T$ and $l(T)=(f/g)_t/g$, we get $(f/g)_t=\lambda_2 g(t)(\varepsilon T+\kappa)^\frac{\rho-4}{2}$.
Finally,
\[
f(t)=
\begin{cases}
\lambda_2g(t)\left(\frac2{\varepsilon(\rho-2)}(\varepsilon T+\kappa)^{\frac{\rho-2}2}+\lambda_3\right),&\mbox{if}\quad \rho\neq2, \\
\lambda_2g(t)\left(\frac1\varepsilon\ln(\varepsilon T+\kappa)+\lambda_3\right),&\mbox{if}\quad \rho=2.
\end{cases}
\]
After re-denoting the constants $\lambda_i$, $i=1,2,3,$ it is easy to see that we get Cases 1 and 2 of Table~2, respectively.
To obtain the corresponding Lie symmetry operators one should make the change of variables
$\tilde t=T,$ $\tilde x=x,$ $\tilde u=u+\mu_2(\varepsilon T+\kappa)^{\frac{\rho-2}{2}}+\mu_3$
(resp. $\tilde u=u+\mu_2\ln(\varepsilon T+\kappa)+\mu_3$ for the second case)
in the vector fields $X_1=\partial_{\tilde x}$ and
$X_2=2(\varepsilon\tilde t+\kappa)\partial_{\tilde t}+\varepsilon\rho\tilde x\partial_{\tilde x}+\varepsilon(\rho-2)\tilde u\partial_{\tilde u}$.

It is interesting to note that the images of two distinct inequivalent cases of Lie symmetry extensions in class (1) (Cases~1 and 2 of Table 2)
belong to the same case of Lie symmetry extensions for class (10) (Case 1 of Table 3).

The other cases are easily treated in the same way.

\section{Conservation laws}
\label{SectionOnCLsOfvcBBMEqs}

We classify (local) conservation laws of equations from class~\eqref{bbm_fgh},
applying the most direct method based on the definition of conservation laws.

We briefly present necessary notions, specifying them to the relevant case of
a single partial differential equation~$\mathcal L$: $L=0$ with the two independent variables~$t$, $x$ and the single dependent variable~$u$,
where $L$ is a differential function, i.e., a smooth function of $t$, $x$ and derivatives of~$u$.
A tuple $(F,G)$ of differential functions is a conserved current of~$\mathcal L$
if $D_tF+D_xG=0$ for any solution of~$\mathcal L$,
where $D_t$ and $D_x$ are the operators of total derivative with respect to $t$ and $x$, respectively.
Then the functions $F$ and $G$ are called the {\em density} and the {\em flux} of the conserved current~$(F,G)$.
Conserved currents $(F,G)$ and $(F',G')$ are equivalent if there exist such differential functions~$\hat F$, $\hat G$ and~$H$
that $\hat F$ and~$\hat G$ vanish for all solutions of~$\mathcal L$~and $F'=F+\hat F+D_xH$, $G'=G+\hat G-D_tH$.
Elements of the factor-set of conserved currents with respect to this equivalence relation are called \emph{conservation laws} of~$\mathcal L$.
Each conservation law~$\mathcal F$ of~$\mathcal L$ admits the representation $D_tF+D_xG=\lambda L$
for some conserved current $(F,G)$ containing in~$\mathcal F$ and a differential function~$\lambda$,
which is called a \emph{characteristic} of the conservation law~$\mathcal F$.
Characteristics~$\lambda$ and~$\tilde\lambda$ are equivalent if their difference vanishes on solutions of~$\mathcal L$.
Under certain natural condition on~$\mathcal L$, there is one-to-one correspondence between
conservation laws and equivalence classes of characteristics~\cite{Olver1993}.

Computing conservation laws for equations from class~\eqref{bbm_fgh},
 we assume that components of conserved currents are of the general form
\[
F=F(t,x,u,u_t,u_x),\quad G=G(t,x,u,u_t,u_x,u_{tt},u_{tx},u_{xx}).
\]
Following~\cite{duzh1984a}, it is possible to show that
in fact any conservation law of each equation from class~\eqref{bbm_fgh} possesses
a~characteristic of order not greater than two
and, therefore, contains a~conserved current whose density and flux orders are not greater than one and two, respectively.
The proof of this is quite long and will be a subject of another paper.

The classification of local conservation laws of equations from the class~\eqref{bbm_fgh} is as follows.

{\it Case 0.}
Each equation from the class~\eqref{bbm_fgh} admits the ``natural'' conservation law with the constant characteristic $\lambda^1=1$.
The corresponding density and flux are
\[
F^1=u,\quad G^1=f(t)u+\frac12g(t)u^2+h(t)u_{tx}.
\]
For general admitted values of the arbitrary elements~$f$, $g$ and~$h$
the associated space of conservation laws is one-dimensional.

{\it Case 1.}
If the arbitrary elements satisfy the equation $((1/h)_t/g)_t=0,$ and, therefore,
\[h(t)=\left(\rho_1\int g(t){\rm d}t+\rho_2\right)^{-1},\]
where $\rho_1$ and $\rho_2$ are constants with $(\rho_1,\rho_2)\ne(0,0)$,
then the space of conservation laws of the corresponding equation of the form~\eqref{bbm_fgh} is at least two-dimensional.
The second basis conservation law can be chosen to have the following characteristic, density and flux:
\begin{gather*}
\lambda^2=\frac uh-\rho_1\left(x-\int\!f(t){\rm d}t\right),\quad
F^2=\frac{u^2}{2h(t)}-\frac{u_x^2}2-\rho_1\left(x-\int\!f(t){\rm d}t\right)u,\\
G^2=\frac{g(t)}{3h(t)}u^3+\frac{f(t)}{2h(t)}u^2+uu_{tx}+\rho_1h(t)u_t-\rho_1\left(x-\int\!f(t){\rm d}t\right)\left(f(t)u+\frac12g(t)u^2+h(t)u_{tx}\right),
\end{gather*}
where $(1/h)_t/g=\rho_1={\rm const}.$

Using the family of point transformations $\tilde t=\rho_1\int g(t){\rm d}t+\rho_2$, $\tilde x=x$, $\tilde u=u/\rho_1$
related to the group~$G^\sim_1$,
we can reduce any equation of this case
with $\rho_1\neq 0$ to the form $u_t+f(t)u_x+uu_x+t^{-1} u_{txx}=0$ (tildes are omitted in the latter equation).

{\it Case 2.}
One more case with at least two-dimensional spaces of conservation laws is given by
the arbitrary elements satisfying the condition $((f/g)_t/g)_t=0,$ i.e., if \[f(t)=\left(\sigma_1\int g(t){\rm d}t+\sigma_2\right)g(t),\]
where $\sigma_1$ and $\sigma_2$ are arbitrary constants.
The second basis conservation law can be chosen to have the following characteristic, density and flux:
\begin{gather*}
\lambda^3=W^2-2\sigma_1  x+2\frac hgu_{tx},\quad F^3=\frac13W^3-2\sigma_1xu-\frac13\frac{f(t)^3}{g(t)^3},\\
G^3=\frac {h(t)}{g(t)} W_t^2+\frac{h(t)^2}{g(t)}{u_{tx}^2}+h(t)W^2u_{tx}-2\sigma_1h(t)xu_{tx}-2\sigma_1f(t)xu-\sigma_1g(t)xu^2+\frac {g(t)}4W^4,
\end{gather*}
where $W=u+f(t)/g(t)$, $(f/g)_t/g=\sigma_1={\rm const}.$

Using the family of point transformations $\tilde t=\sigma_1\int g(t){\rm d}t+\sigma_2$, $\tilde x=\sigma_1 x$, $\tilde u=u$
related to the group~$G^\sim_1$,
we can reduce any equation of this case with $\sigma_1\neq 0$ to the form $u_t+t u_x+uu_x+h(t) u_{txx}=0$ (tildes are omitted in the latter equation).

{\it Case 3.}
The maximal dimension of the spaces of conservation laws for equations from class~\eqref{bbm_fgh} equals three
and is reached for the intersection of Cases~1 and~2,
where arbitrary elements satisfy the both constraints, $((f/g)_t/g)_t=0$ and $((1/h)_t/g)_t=0$.
Then for each of the spaces, a basis consists of conservation laws with the characteristics $\lambda^1$, $\lambda^2$ and~$\lambda^3$ and
the conserved currents $(F^1,G^1)$, $(F^2,G^2)$ and $(F^3,G^3)$, respectively.
The corresponding equation can be reduced to the form
$u_t+(\sigma_1t+\sigma_2)u_x+uu_x+(\rho_1t+\rho_2)^{-1}u_{txx}=0$ by transformation~\eqref{tr}.
The further simplification is possible by transformations from the group~$G^\sim_2$.
For example, we can set one of the linear combinations $\sigma_1t+\sigma_2$ or $\rho_1t+\rho_2$ to $t$ if $\sigma_1\neq0$ or $\rho_1\neq0$, respectively.

A well-studied subcase of Case~3 is constituted by constant-coefficient equations, for which $\rho_1=\sigma_1=0$
\cite{benj1972a,duzh1984a,Olver1979}.
\noprint{
Each equation of the form~\eqref{bbm_fgh} with constant $f$, $g$ and $h$
have three linearly independent conservation laws with characteristics $1$, $u$ and $(u+f/g)^2+2h u_{tx}/g$.
The corresponding densities and fluxes are
\begin{gather*}
\tilde F^1=u,\quad \tilde G^1=fu+\frac g2u^2+h u_{tx},\\
\tilde F^2=\frac{u^2}2-h\frac{u_x^2}2,\quad \tilde G^2=\frac f2u^2+\frac g3u^3+h uu_{tx},\\
\tilde F^3=\frac{1}3\left(u+\frac fg\right)^3,\quad \tilde G^3=\frac hg u_t^2+\frac{h^2}gu_{tx}^2+h \left(u+\frac fg\right)^2u_{tx}+\frac g4 \left(u+\frac fg\right)^4,
\end{gather*}
}
Up to $G^\sim_1$-equivalence any constant-coefficient equation from class~\eqref{bbm_fgh} can be mapped to the equation
$u_t=uu_x+\varepsilon u_{xxt},$ where $\varepsilon=\mathop{\rm sgn}h=\pm1$.
Then the characteristics and the components of the conserved currents of the above basis conservation laws take the form (cf.~\cite[p~195]{Ibr})
\begin{gather*}
\lambda^1=1,\quad F^1=u,\quad G^1=-\frac12u^2-\varepsilon  u_{tx},\\
\lambda^2=\varepsilon u,\quad F^2=\frac{u^2}2+\varepsilon \frac{u_x^2}2,\quad G^2=-\frac 13u^3-\varepsilon uu_{tx},\\
\lambda^3=u^2+2\varepsilon u_{tx},\quad F^3=\frac13u^3,\quad G^3=\varepsilon u_t^2-u_{tx}^2-\varepsilon u^2u_{tx}-\frac14 u^4.
\end{gather*}

\section{Conclusion}

The aim of the present paper is to enhance and generalize existing results on Lie symmetries and conservation laws
of variable-coefficient BBM equations of the form~\eqref{bbm_fgh}.
Comparing  the results of~\cite{Molati&Khalique2013} with those collected in Table~2,
we conclude that Lie symmetry extensions presented in~\cite{Molati&Khalique2013}
are particular specifications of Cases~1--5 from Table~2 for certain fixed values of the arbitrary element~$g$.
For example, there are two cases in the classification list derived in~\cite{Molati&Khalique2013}
with the three-dimensional maximal Lie symmetry algebras (Cases~4 and~10 of Table~1 therein).
These cases are particular subcases of Case~5 of Table~2 for $g=g_0={\rm const}$ and $g=g_0e^{kt}$.

The results of Section~\ref{SectionOnCLsOfvcBBMEqs} on conservation laws are quite expectable
and, at the same time, are not trivial.
They naturally generalize well-known results of constant-coefficient BBM equations
and need the completion of the most significant and tricky part of the proof,
which is deriving an upper bound for order of conservation laws similarly to \cite{duzh1984a}.

\ack{This work and the participation of OV in the Workshop was partially supported by a grant from the Niels Henrik Abel Board.
OV would like to thank the University of Cyprus for hosting her during this project.
The research of ROP was supported by the Austrian Science Fund (FWF), project P25064.
ROP is also grateful for the hospitality and financial support provided by the University of Cyprus.}

\expandafter\ifx\csname url\endcsname\relax
  \def\url#1{{\tt #1}}\fi
\expandafter\ifx\csname urlprefix\endcsname\relax\def\urlprefix{URL }\fi
\providecommand{\eprint}[2][]{\url{#2}}

\end{document}